# Commissioning of the Low Energy Electron Gun Test Stand at the University of Chicago*

S. Kladov[†], M. Bossard[1], J. Brandt[1], Y-K. Kim[1], The University of Chicago, Illinois, USA
N. Banerjee[2], G. Stancari[2], B. Cathey[2], Fermilab, Batavia, Illinois, USA
S. Nagaitsev[3], Thomas Jefferson National Accelerator Facility, Newport News, Virginia, USA

*Abstract*

We built a test stand for evaluating the performance of the thermionic electron sources for the electron lens project at the Integrable Optics Test Accelerator (IOTA) in Fermilab. The lens will be used to study nonlinear dynamics and electron cooling of 2.5 MeV protons with strong space charge. The test stand will validate the characteristics of the thermionic sources and the main parameters of the generated beams. In this paper we present the results of the commissioning of the UChicago test stand and validation of the hollow beam source.

## INTRODUCTION

The Integrable Optics Test Accelerator (IOTA) is an easily re-configurable 40 m storage ring at Fermilab. It is dedicated to accelerator research, including nonlinear dynamics and intense beams with strong space-charge. Flexible lattice allows it to circulate as electrons, as protons, at kinetic energies up to 150 MeV and 2.5 MeV, respectively. [1,2]

A new upgrade of the IOTA ring involves an **electron lens**, that is an electron beam, co-propagating with circulating protons. The lens can be used for electron cooling, collective effects compensation, to study nonlinear dynamics and more. The electron lens was successfully implemented on various rings, such as RHIC and Fermilab Tevatron [3,4].

The IOTA layout, as well as the path of the electron beam, is shown on Fig. 1. The proposed electron lens scheme for IOTA is presented on Fig. 2. [5]

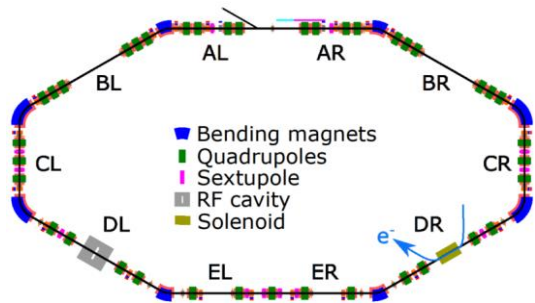

Figure 1: IOTA ring layout. The lens will be installed in the DR line with zero dispersion.

___________________
* Work supported by The University of Chicago and Fermi National Accelerator Laboratory
† kladov@uchicago.edu

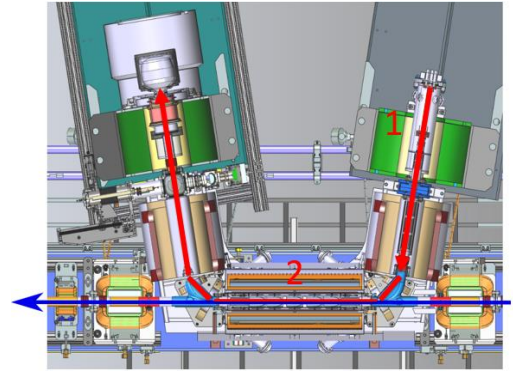

Figure 2: Electron lens scheme. Electrons, produced in a thermionic source (1), travel with the circulating proton beam (2).

Our goal for this project is to construct a test stand at the University of Chicago that meets the specific parameters required by IOTA ring at Fermilab. The test stand will serve as a crucial step in ensuring the optimal performance of the thermionic electron sources for the electron lens project.

This paper describes the commissioning of the test stand, which utilizes a hollow thermionic source originally used for collimation in the Tevatron.[4] The commissioning process includes testing the filament, vacuum, control system, and more. In this paper, we focus on the main beam parameter validation, specifically the beam's transverse distribution and current.

First we describe the test stand and its operation parameters. Then we present the experimental results: magnetic field verification, beam profile and current measurements. Finally, we summarize our work and present future plans.

## TEST STAND

Figure 3 shows the construction of the test stand for validating thermionic sources. While the design has undergone several changes, the main component is a diagnostic cube (labelled 2 in Fig. 3) placed below a long spool piece which houses the thermionic electron source under test placed in the field of a solenoid. The diagnostic cube has two quartz windows through which we can image the beam profile and cathode itself. The entire setup is maintained under high vacuum conditions between $10^{-9}$ mbar up to $10^{-7}$ mbar during source testing. Table 1 shows the range of the main parameters in the test stand.

The test stand's diagnostics measure vacuum, cathode temperature, electron beam current, and profile. Vacuum is measured using a hot filament ion gauge and a Residual Gas Analyser (labelled 5 in Figure 3). Pressure is maintained using a turbo molecular pump and two ion pumps.

Cathode temperature is optically measured using ratio pyrometry, with a CMOS camera located below a window in the bottom of the diagnostic cube.

Table 1: Main test stand parameters

| Parameter | Range | Units |
|---|---|---|
| Solenoid field | 0-250 | G |
| Cathode voltage | 0-5000 | V |
| Cathode current | 0-3 | A |
| Pressure | $10^{-9}$-$10^{-7}$ | mbar |
| HV pulse length | 1-1000 | μs |

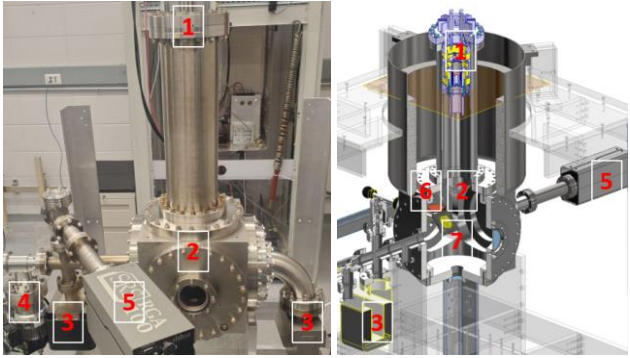

Figure 3: Test stand photo and CAD model. 1. Thermionic source, 2. Diagnostics cube, 3. Ion pumps, 4. Turbo molecular pump, 5. RGA, 6. Faraday cup, 7. YAG screen

The main beam parameters: transverse profile, and current, are measured inside the diagnostics cube using a Faraday cup and a scintillating screen. The Faraday cup collects electrons from the beam, and the resulting current measured in galvanically isolated way using an AC current transformer. Secondary electrons generated by primary electrons striking the copper Faraday cup can backscatter and affect current measurements. A biasing voltage of 50-100 V suppresses secondary electrons. A Cerium doped Yttrium Aluminum Garnet (Ce:YAG) crystal acts as a scintillator that produces green light with intensity proportional to the beam current incident on its surface. The scintillation light is imaged through a quartz window to the side of the diagnostic cube using a CMOS camera.

The high voltage is pulsed in order to keep the pressure low enough for the normal cathode operation, and also to mimic the actual source behavior, where the electron beams should be generated on each proton revolution. Usual pulse length used in the commissioning is 10 μs, while the minimum length can be lower than 1 μs.

## EXPERIMENTAL RESULTS

### Magnetic field verification

Magnetic field of the test stand solenoid is not strong enough to perfectly confine the electron beam. As a result, particles' cyclotron radius is not negligible, and the amount of spiral revolutions that particles take during the motion in solenoid is not large enough. This leads to **scalloping** – periodic behavior of the beam distribution as a function of the beam energy. The exact distribution predictions in the presence of the space charge cannot be easily made without simulations. However, it is easy to obtain theoretical scalloping period dependence on the beam energy.

The number of revolutions $N$, a particle of energy $E$ completes inside a solenoid of length $h$ and uniform axial field $B$ is given by

$$N(E) = \frac{Be}{m}\frac{1}{2\pi}\frac{h}{c} \, 1/\sqrt{1 - \left(\frac{E_0}{E_0+E}\right)^2}, \quad (1)$$

where $E_0 = 511$ keV is the electron rest mass, $c$ is the speed of light and $v$ is the longitudinal velocity of the electron as it passes through the solenoid.

Difference in one cyclotron revolution particles take throughout the solenoid corresponds to one period of scalloping, i.e., $N(E_1) - N(E_2) = 1$. The equation can be solved for $T(E_1)$, where $T = E_2 - E_1$, that can be measured experimentally.

The electron beam from the hollow source is larger than the Faraday cup, so the measured current strongly depends on the beam size. Scalloping results in the fluctuations of this size, that we can see in the current dependence on the beam energy. This dependence, measured with 50 eV step, is shown on the Fig. 4 (left). Note that the current here is not normalized because we don't need the exact value to see the scalloping. Figure 4 (right) shows the comparison of the theoretical predictions and experimental results. The grey area corresponds to the uncertainty of magnetic field measurement inside the solenoid. The measured scalloping periods agree very well with the measured magnetic field, thus verifying the uniformity of the field seen by the electron beam inside the solenoid.

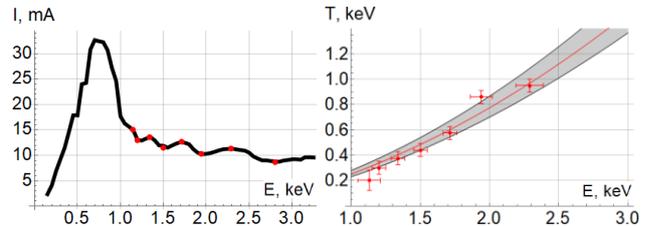

Figure 4: Left: measured beam current dependence on the beam energy. Red dots mark the points chosen for $E_1$; Right: Dependence of scalloping period $T$ on $E_1$. Comparison of the theoretical prediction and measurements.

### Beam Distribution

The beam distribution is imaged using a Ce:YAG scintillating screen with an usable diameter of 46 mm placed at an angle of 45° with respect to the direction of the electron beam. Hence the electrons effectively see an ellipse with major and minor axes sizes of 46 mm and 32 mm respectively. The hollow source used for commissioning generates a beam with an outer diameter of 24 mm which should fit in the screen. However, due to insufficient magnetic focussing, the beam diverges at the position of the Ce:YAG screen. It still preserves an axially symmetrical shape, that allows us to measure a part of the beam profile by intentionally deflecting the beam out of the YAG screen centre.

Although we need to test only the desired beam energy, it is helpful to be able to capture the beam even on low voltages. However, the light emitted by the scintillator is superposed on top of a reflection of the image of a hot glowing cathode and other scattering light. To improve the signal-to-noise ratio of the beam profile, we employ a subtraction technique as follows. Fixing the cathode voltage and hence the current, we adjust the duty cycle of the electron beam. The total illumination of the scintillator is proportional to the average time the beam is generated. We then perform a linear regression of the green colour by the duty cycle frequency. Only the scintillation caused by the beam is visible on these pictures.

The dependence of the axially symmetrical beam density on the radius is measured by averaging one arch of beam imaged discussed above. The centre of the distribution is reconstructed from the shape of the arch. An example of the beam arch picture at 1400 eV is presented on Fig. 5 (left). The scalloping becomes apparent on the profile pictures, one of which is shown on Fig. 5 (right).

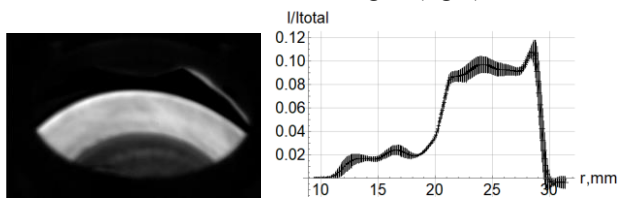

Figure 5: Part of the beam distribution at 1400 eV and the corresponding beam profile.

The profile is compared with 2d TRACK simulations [6]. Although the beam resembles the predicted distribution, the dimensions are larger in approximately 2-3 times due to the insufficient magnetic field.

*Beam Current*

The Faraday cup is almost the same size as the Ce:YAG screen and is oriented perpendicular to the path of the electrons. However, since the beam from the hollow source diverges, only a fraction of the total beam current is incident on the Faraday cup.

The response of the measurement network to the rising edge of the current pulse seen by the Faraday cup includes a resonant ringing along with an exponentially decay when the electron current is constant. Both the height of the peak of the response and the total integrated value of the detected pulse are proportional to the current, with a calibration factor provided by the manufacturer.

The beam current dependence on the cathode voltage should be given by the Child's law ($I = \rho U^{3/2}$, where $\rho$ is a perveance of the source, depending on its geometry), that is our first confined-beam commissioning check. This dependence for different (intentionally large for the purposes of testing) biasing voltages, as well as the 3/2 law fit, is shown on Fig. 6 (left). The beam gets wider after ~800 V, and the 3/2 law is not fulfilled there.

The TRACK-predicted current for 800 V is approximately 130 mA, that is much higher than obtained experimentally. We can estimate the actual beam current by dividing the measured current by the fraction of the beam that lands on the Faraday cup, which we can estimate from the beam profile measurement. Such reconstruction results are presented in Fig. 6 (right). Even though this method yields large uncertainties, the simulation curve is consistent with the calculated current from measurements.

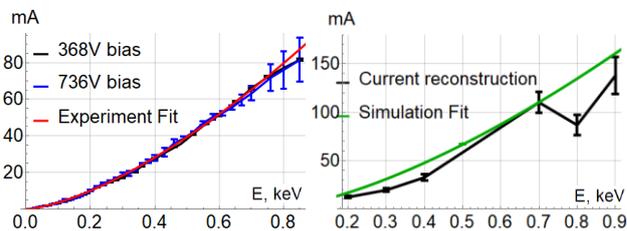

Figure 6: Measured beam current with different FC biasing voltage and the corresponding current reconstruction.

## CONCLUSION

The test stand was tested on the hollow gun the beam distribution and the beam current were confirmed with simulations and expectations. The magnetic field seen by the electron beam was verified by measuring the scalloping period and it matched well with hall probe measurements of the solenoid field.

As the result. the UChicago test stand for the low energy thermionic sources is in fully working condition, and the commissioning is almost finished. Cathode temperature automatic measurements and some control system improvements will be done in the nearest future.

The IOTA electron lens project is, on the contrary, only starting. Various cathode configurations will be tested on the test stand, and the lens will be built based on the results of these experiments.

## ACKNOWLEDGEMENTS

This manuscript has been authored by Fermi Research Alliance, LLC under Contract No. DE-AC02-07CH11359 with the U.S. Department of Energy, Office of Science, Office of High Energy Physics. The project is also supported by the University of Chicago.